\documentclass{PoS}
\usepackage{amssymb, times, mathptmx, graphicx}

\title{Revealing the Nature and Location of High Energy Emission in the Candidate Binary SMBH System OJ 287}

\ShortTitle{Short Title for header}

\author{\speaker{Pankaj Kushwaha}, E. M. de Gouveia Dal Pino\\
         Department of Astronomy (IAG-USP), University of Sao Paulo, Sao Paulo 05508-090, Brazil\\
        E-mail: \email{pankaj.kushwaha@iag.usp.br}}

\author{Alok C. Gupta\\
       Aryabhatta Research Institute of Observational Sciences (ARIES),Manora Peak, Nainital 263002, India}
\author{Paul J. Wiita\\
       Department of Physics, The College of New Jersey, P.O. Box 7718, Ewing, NJ 08628-0718, USA}

\abstract{The latest flare of the regular $\sim$ 12 years quasi-periodic optical outbursts
in the binary SMBH candidate system OJ 287 occurred in December 2015. Following this,
the source has exhibited enhanced multi-wavelength (MW) variability in spectral,
temporal and polarization domains with new features never seen before. Our MW
investigation show that the overall MW variability can be divided into two-phase,
(i) November 2015 -- May 2016 with variability from near-infrared (NIR) to Fermi-LAT
$\rm \gamma$-ray energies (0.1 -- 300 GeV), and (ii) September 2016 -- July 2017 with intense NIR
to X-ray variability but without any activity in the Fermi-LAT band, and the very
first detection at very high energies (VHEs, E $>$ 100 GeV) by VERITAS. The broadband
SEDs during the first phase show a thermal bump in the NIR-optical region and a
hardening in the $\rm \gamma$-ray spectra with a shift in its peak. The thermal bump like
feature is consistent with the description of the standard accretion-disk associated
with the primary SMBH of mass $\sim 1.8\times10^{10} M_\odot$ while the $\rm \gamma$-ray emission
can be naturally reproduced by inverse Compton scattering of photons from the broad
line region which has been seen during the close encounter duration of the binary
SMBHs, thereby suggesting a sub-parsec scale origin. The SEDs during the second
phase (VHE detection) is a mixture of typical OJ 287 SED with hardened $\rm \gamma$-ray spectra
and an HBL SED and can be explained in a two-zone model, one located at sub-parsec
scales and other at parsec scales. During both the phases, the MW variability is
simultaneous and almost always accompanied by changes in the polarization properties,
exhibiting random and systematic variations, suggesting a strong role of magnetic
field and turbulence.}

\FullConference{International Conference on Black Holes as Cosmic Batteries: UHECRs and Multimessenger Astronomy - BHCB2018\\
		12-15 September, 2018\\
		Foz du Iguazu, Brasil}

\begin{document}

\section{Introduction}
OJ 287 belongs to the BL Lac (BLL) subclass of blazars -- active galactic nuclei with
relativistic jet directed at close angles to our line of sight and characterized
by featureless emission spectra or a weak signal of optical emission lines
superimposed on it. Identified in 1967 \cite{1967AJ.....72..757D} and located at
the redshift of $\rm z~=~0.306$ \cite{1985PASP...97.1158S,1989A&AS...80..103S},
its radio and optical brightness combined with a diverse range of peculiar properties
have made it one of the best-studied sources across the entire electromagnetic (EM)
spectrum \cite[and references therein]{2018MNRAS.479.1672K,2018MNRAS.473.1145K}
and a potential target for many of the unresolved problems in relativistic jet
physics \cite{2018MNRAS.478.3199B,2018MNRAS.473.1145K,2013MNRAS.433.2380K} as well as
general relativity \cite[and references therein]{2016ApJ...819L..37V}. The source
central engine has been argued to be a precessing binary supermassive black hole
(SMBH) system  of masses $1.8\times10^{10}~M_\odot$ and $1.5\times10^{8}~M_\odot$
from an $\sim 11.65$-year periodicity in the optical light curve \cite[and references
therein]{2016ApJ...819L..37V}. Though the model has undergone many iterations
from its first inception, the basic tenets and claim remain the same. It predicted
the next outburst between the end of 2015 and the beginning of 2016 depending on 
the spin of the primary SMBH. Contrary to this, radio kinematic studies suggest
$\sim~22$-year periodicity and rather argue and favor a jet-disk central engine
with jet rotating around its axis and precessing as well \cite[and references
therein]{2018MNRAS.478.3199B}.

One of the fundamental problems in the blazars' research is the process/mechanism responsible
for the high-energy hump in their characteristic broad double-humped spectral energy
distributions (SEDs). Broadly two possible channels: leptons and hadrons are the
primary candidates. In the former, the high-energy hump originates as a result
of inverse Compton (IC) scattering of the local soft photons fields while it is due
to proton-proton and/or proton-photons initiated cascade processes in the case of
the latter. The widely considered target photon fields for the leptonic channel
includes the synchrotron photons within the jet and accretion disk (AD), broad
line region (BLR), IR torus (IR), and cosmic microwave background photons (CMB)
external to the jet. The IC scattering of synchrotron photons is called synchrotron
self-Compton (SSC) while IC of an external field is generally called external Comptonization
(EC) with EC-AD, EC-BLR, EC-IR, and EC-CMB representing the respective IC spectrum
of the mentioned external fields. The low-energy component of the SED which spans radio to ultraviolet (UV)
and even to X-ray energies is widely regarded as synchrotron emission from relativistic
non-thermal electrons in the jet due to its high and variable polarization. Temporally,
flux variability is stochastic, present on all time-scales from minutes to decades
with statistical properties broadly consistent with other accretion-powered
sources \cite{2017ApJ...849..138K,2016ApJ...822L..13K}, and often accompanied
by changes in spectral and polarization properties.

The presence of synchrotron emitting relativistic electrons makes IC process as one
of the natural candidates for the origin of the high-energy component of the blazars' SEDs.
In this scenario, the high-energy component of BLLs is widely attributed to the
SSC process due to the lack of strong photon fields external to the jet. The BLL 
object OJ 287, from the perspective of broadband SED, belongs to the low-frequency peaked BLL with the
low-energy component peak at or below NIR bands and the high-energy peak at 100
MeV or below \cite{2010ApJ...716...30A,2013MNRAS.433.2380K}. Detailed modeling of
source's broadband SEDs by \cite{2013MNRAS.433.2380K} during one of the best
observed multi-wavelength (MW) flares in 2009 showed that observational constraints
from the different observed flux states rules out the SSC process for MeV-GeV $\gamma$-ray
origin. Rather it requires
both SSC (for X-ray) and EC-IR from a torus of temperature of $\sim 250$ K to
explain the second energy component \cite{2013MNRAS.433.2380K}.

In this proceeding, we report the results of MW spectral and temporal analysis of
the two of the outburst phases shown by OJ 287 between end 2015 to mid-2017 as 
reported in \cite{2018MNRAS.479.1672K,2018MNRAS.473.1145K}. 

\section{Multi-wavelength data}
\begin{figure}
 \centering
 \includegraphics[scale=1]{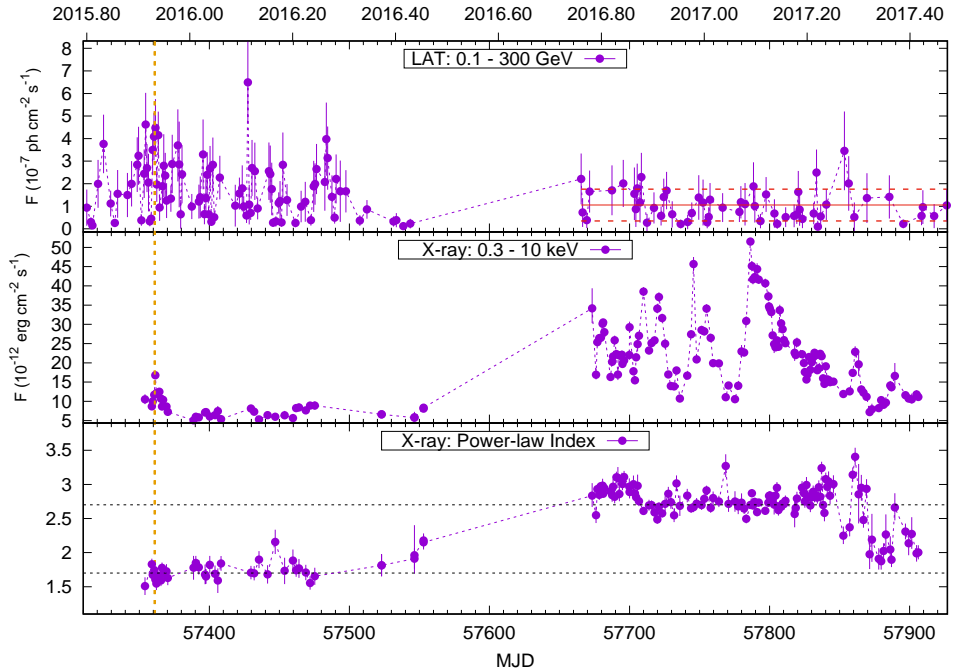}
 \caption{The $\gamma$-ray and X-ray light curves of OJ 287 between 15 October 2015
 to 26 June 2017 along with the corresponding X-ray power-law index. The $\gamma$-ray points are extracted
 from \emph{Fermi}-LAT data binned over 1-day while the X-ray data correspond to the different
 observation IDs of \emph{Swift}-XRT. The NIR-optical flux variability trend is similar
 to the X-ray \cite[for more details]{2018MNRAS.479.1672K,2018MNRAS.473.1145K}. The
 vertical line marks the impact outburst \cite{2016ApJ...819L..37V} and solid red
 horizontal line in the top panel
 represent the flux mean over the plotted duration while the dotted lines show the
 mean flux-error around it.}
 \label{fig:HElc}
\end{figure}

Since the outburst was anticipated as per the binary SMBH model, OJ 287 was under
monitoring and was followed by a large number of observatories in a coordinated
fashion as soon as the first sign of an increase in activity was observed
\cite{2016ApJ...819L..37V,2018arXiv180303964G}. A typical trend of multi-band flux
variability observed following the first sign of increased activity till mid-2017 (15
October 2015 -- 26 June 2017) is shown in figure \ref{fig:HElc} by the $\gamma$-ray and
X-ray light curve from \emph{Fermi}-LAT and \emph{Swift}-XRT respectively. The
flux variability trend in NIR to ultraviolet (UV) bands was similar to the X-rays.
The full MW light curves and the details of observations are presented in the
\cite{2018arXiv180303964G,2018MNRAS.479.1672K,2018MNRAS.473.1145K,2017MNRAS.465.4423G}.

\section{Discussion}
The rise of NIR-optical emission arrived around the time predicted by the binary
SMBH model and thus was followed across the EM spectrum in a coordinated way by
different groups and observatories \cite{2016ApJ...819L..37V,2017MNRAS.465.4423G}.
Since then till mid-2017, OJ 287 has shown two phases of high MW activity, both
characteristically different from each other as well as with all the previously reported
short duration MW activity in terms of flux variability and the polarization
properties. The first phase corresponds to the activity seen during and after the
impact outburst till the source flux almost attained the pre-outburst level, roughly
November 2015 -- May 2016 (MJD 57315 -- 57460) period \cite{2017MNRAS.465.4423G,
2018MNRAS.473.1145K}. The outburst at MJD 57361, marked by the vertical line in
figure \ref{fig:HElc} is the claimed impact outburst resulting from the impact
of secondary SMBH on the accretion disk of the primary \cite{2016ApJ...819L..37V}.
It has a relatively low degree of polarization (< 10\%) but marks the beginning of
a huge systematic swing of polarization angle by $\sim~200^\circ$ \cite{2017MNRAS.465.4423G}.
It continued to be observed in the optical even after it reached its pre-outburst flux level,
but only in a few bands \cite{2017MNRAS.465.4423G}, until the yearly gap when it becomes too
close to the Sun. As soon as observations were resumed following the yearly gap,
it was already in a high NIR-optical flux states and  thus, was observed across
the EM bands \cite{2018arXiv180303964G,2018MNRAS.479.1672K}. Observations in X-ray
showed it to be in a historic high state and thus was observed by the VERITAS under
the target of Opportunity source which resulted in its first-ever detection at VHEs
\cite{2017arXiv170802160O}.

In terms of spectral and temporal MW trends so far reported in the literature over
a short duration, the current two phases are unique in many ways. Temporally, the 
MW flux variations seen during the first phase are similar to previously reported MW
activity during the flares with flux variations from NIR to MeV-GeV $\gamma$-rays
and all being simultaneous within their respective observational cadence
\cite{2018MNRAS.473.1145K,2013MNRAS.433.2380K}. The variations are associated with
frequent and strong changes in optical PA and PD. On the contrary, the MW flux
variability during the second phase displays an unreported combination
of observations where OJ 287 showed intense NIR
to X-ray variation but statistically no variability in MeV-GeV energy band though
it is detected quite frequently over daily binning of the data. During 
this phase too, the NIR to X-ray variations are simultaneous for all except towards
the end duration, during which it attained its lowest flux level, similar to the
low-level flux of the first phase. This period instead suggests a hint of systematic
variation with flux rise appearing first at high energies and then subsequently at
lower energies. The polarization variations are also quite different with PD variation
being like that in the first phase, while the PA changed rather smoothly and systematically
by $\sim~100^\circ$ with fluctuations superposed on it over the duration of the second
phase \cite{2018MNRAS.479.1672K}.  

\begin{figure}
 \centering
 \includegraphics[scale=1]{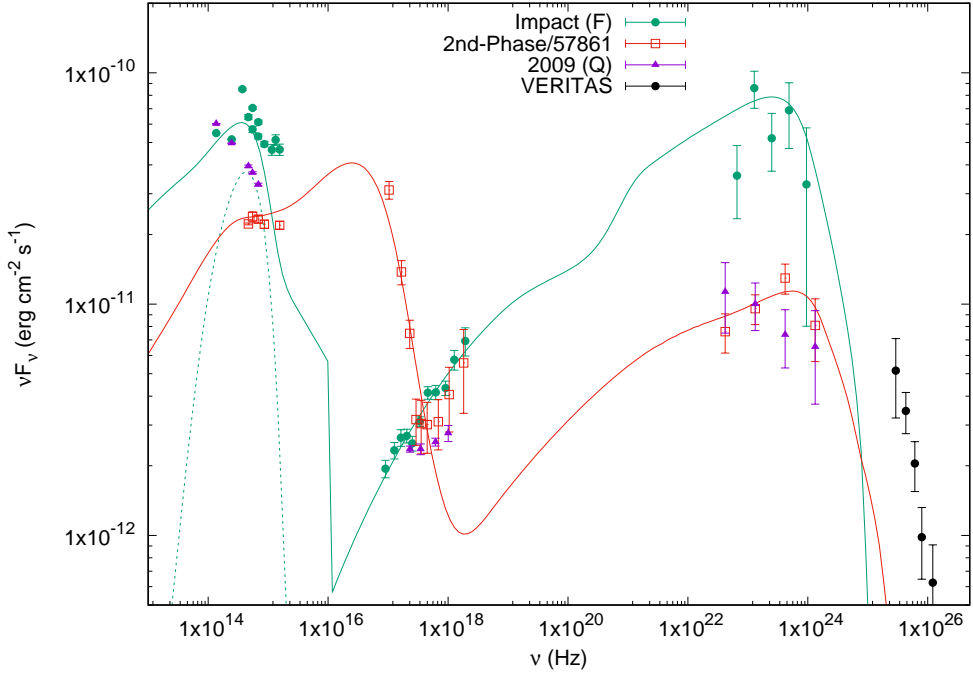}
 \caption{Broadband SEDs of OJ 287 corresponding to different flux states (Flare: F,
 quiescent: Q) of the source.}
 \label{fig:seds}
\end{figure}

From energy spectral point of view too, the two duration are completely different with
respect to each other (see figure \ref{fig:seds}) as well as with all the previously
reported broadband SEDs of
the source in literature till date \cite{2013MNRAS.433.2380K,2010ApJ...716...30A}.
For the first phase, the broadband SED constructed for the impact flare which has
a good coverage from NIR to MeV-GeV $\gamma$-rays shows a break in the NIR-optical
spectrum and also a hardening of the MeV-GeV spectrum with a shift in its peak to higher
energies \cite{2018MNRAS.473.1145K}. The MeV-GeV spectral hardening can also be
seen in the broadband SED constructed for the low flux state which, unfortunately,
lacks the NIR-optical data. Either of the broadband SED, however, show no change
in the X-ray spectrum. Focusing on the spectral shape of NIR data (J and K bands)
of impact SED clearly show that there is no shift in the location of NIR-optical
peak with respect to the 2009 NIR spectra and thus suggest that the NIR-optical
break is likely an additional emission component. On the other hand,
in the leptonic emission scenario, the spectral shape and the location of SED peaks
of the low-energy component is related to the corresponding spectral shape and the
location of peak of the high-energy component. Thus, the lack of change in the NIR-optical
spectrum and its peak location rules out the previous EC-IR of a $\sim 250$ K 
torus interpretation \cite{2013MNRAS.433.2380K} for the  current MeV-GeV spectral
hardening and peak shift. Combining the no shift and change in the NIR part
with no change in X-rays suggests the MeV-GeV emission to arises from a different process
or IC of photon field. Surprisingly, line emission in OJ 287 has been detected
during the impact duration of 1984 and 2005 -- 2008 \cite{2010A&A...516A..60N}
and the first phase is one such. Addition of a BLR-IC component naturally reproduces
both the MeV-GeV spectral hardening as well as the shift in its peak while at
the same time is consistent with the NIR shape and its peak location. This, in
addition, indirectly constrains the location of the emission region to be within
BLR region which is on sub-parsec scales, contrary to the previous claim of
parsec-scale location \cite{2013MNRAS.433.2380K}. On the other
hand, the NIR-optical break is naturally produced by the accretion-disk 
emission associated with the primary SMBH of mass $\sim~1.8\times10^{10}~M_\odot$
\cite{2018MNRAS.473.1145K}. Surprisingly, its first appearance in end May 2013 (MJD 56439)
is temporally coincident with the time of impact of the secondary in the SMBH
frame as predicted by the binary SMBH model \cite{2016ApJ...819L..37V}, thereby
favoring a binary SMBH central engine over other claims/interpretations
\cite{2018MNRAS.478.3199B}.

As far as the spectral evolution during the second phase is concerned, the broadband
SEDs during the high optical to X-ray flux states are superposition of the typical broadband
SED of OJ 287 with the modified MeV-GeV spectrum and a high-frequency peaked BLL
(HBL) SED while the low flux or quiescent state SED for this duration is similar
to the low flux SED of the previous phase. Further, the LAT spectrum for 
the VHE episode is consistent with the extrapolation of the VERITAS VHE spectrum
to the LAT band \cite{2018MNRAS.479.1672K}. Focusing on the departure of the high-energy
end of X-ray spectrum with respect to its lower-energy end and its level to that
of the quiescent X-ray spectrum from the first phase, one can clearly see that
this new high X-ray emission is nothing but an additional HBL component
with its peak in UV to soft X-ray region (see
fig. \ref{fig:seds}, \cite{2018MNRAS.479.1672K}). Accordingly, the overall SED can be 
modeled with a two-zone model with the first zone responsible for the typical OJ 287
SED with modified MeV-GeV spectrum while the other zone produces an HBL like SED,
resulting from the SSC + EC-IR \cite{2018MNRAS.479.1672K}.

The non-variability of MeV-GeV band can also be clearly understood from the
extracted SEDs during the second phase. Since the photon-flux is based on counting,
the overall integrated photon count over the instrument band is dominated by the
photons at the low-energy end for sources with power-law emission as typically
seen in nature. Focusing over the first data point at MeV-GeV $\gamma$-ray band
which is consistent with the corresponding 2009 quiescent state flux (see fig.
\ref{fig:seds}) while the peak SED point is at most $\sim$ 1.5 times to its 2009
value. This translates at most to a similar increment of photons in the energy band
of the peak point (energy deposited $\propto$ number of photons). Thus, the
order of magnitude difference in the energy of the two bands transforms
to a 4 order of magnitude difference in the respective photon counts assuming
a power-law spectral index of 2 and hence,
no appreciable change in the observed photon flux vis-a-vis 2009 quiescent
photon flux.

To sum up, the MW activity seen during 15 October 2015 -- 26 June 2017 revealed
a completely new spectral and  temporal phase of OJ 287 and resolved the
fundamental problem of nature of MeV-GeV emission in OJ 287 and indirectly provided the
location of emission region. The association of most of the flux changes with PD
and a systematic as well as fluctuating change in PA suggests a strong role
of the magnetic field and turbulence for the observed features. The activity and
its occurrence close to the timing predicted by the binary SMBH model makes OJ 287
as one of the best potential candidates to explore SMBH Jet-disk connection and
tie down high energy emission processes within the central engine.

\acknowledgments
PK and EMDGP acknowledge support from the Brazilian agencies FAPESP (grant 2015/13933-0
and 2013/10559-5) and CNPq (grant 308643/2017-8).

\bibliographystyle{JHEP}
\bibliography{nth_v5.bib}

\end{document}